\documentclass[a4paper]{article}
\usepackage{amsmath,amssymb,mathrsfs}
\usepackage[light,condensed,math]{iwona}
\usepackage[T1]{fontenc}
\usepackage[top=3cm,bottom=3cm,left=3cm,right=3cm]{geometry}
\date{}
\begin{document}
\author{Burak Tevfik Kaynak${^\dag}$ and O. Teoman Turgut$^\ddag$ \\ Department of Physics, Bo\u{g}azi\c{c}i University \\ 34342 Bebek, Istanbul, Turkey \\ $^\dag$burak.kaynak@boun.edu.tr, $^\ddag$turgutte@boun.edu.tr}
\title{\bf A Klein Gordon Particle Captured by Embedded Curves}
\maketitle
\begin{abstract}
In the present work, a Klein Gordon particle with singular interactions supported on embedded curves on Riemannian manifolds is discussed from a more direct and physical perspective, via the heat kernel approach. It is shown that the renormalized problem is well-defined, and the ground state energy is unique and finite. The renormalization group invariance of the model is discussed, and it is observed that the model is asymptotically free. 
\end{abstract}
\section{Introduction}
Schr\"{o}dinger operators with singular interactions have been studied in depth for sometime by now. The related literature is vast, and the reader is invited to consult Refs.~\cite{al1,al2}, and the references therein for extensive studies. On the other hand, similar problems in  which  interactions are supported by curves attracted mathematicians' and mathematical physicists' attention as a result of the works of Refs.~\cite{bra,ex1}. The physical motivation behind those studies stems from the need for modeling semiconductor quantum wires~\cite{ex2}. In these studies, the Schr\"{o}dinger operator with singular interactions supported by curves was introduced as a model for quantum mechanics of electrons confined to narrow tube-like regions. Following these works, the theory of curve supported singular interactions in $\mathbb{R}^3$  from the point of view of self-adjoint extensions is developed rigorously in the works~\cite{ex3,ex4,ex5,ex6,ex7,lob}, and also in the references therein. The authors in those works also study the positivity of the ground state wave function, the regularity of the eigenfunctions, and the case of periodic curves as well from the same point of view. The problem generally admits an infinite number of extension parameters. However it is also natural to choose a single extension parameter for simplicity. As a result, the value of the emerging coupling constant is chosen to define the theory. Therefore, it is natural to investigate the range of parameters which would lead to bound states.

In contrast to the aforementioned studies, the present authors follow a more direct approach in Ref.~\cite{kt}, which is more physical despite being less rigorous, in order to study the Schr\"{o}dinger operator with singular interactions supported by curves embedded in a Riemannian manifold. They choose the bound state energy of the curve as the defining parameter. As the bound state problem is nonperturbative, they construct the resolvent for the full Hamiltonian in terms of the so-called principal operator, as having been done to study the nonrelativistic point interaction on Riemannian manifolds in Ref.~\cite{et} and to study relativistic bosons with delta function potentials on Riemannian manifolds in Ref.~\cite{ct}. The distinctive advantage of this formulation is that once the renormalization of this operator is accomplished, one can easily scrutinize various crucial aspects of the model just by working out that operator, such as, whether the ground state energy is bounded from below or not, the positivity of the ground state, and the renormalization group equation, which the principal operator is expected to obey.

The aim of this work is to generalize the results of Ref.~\cite{kt} to the case of a relativistic particle with singular interactions supported by curves embedded in a Riemannian manifold, and to demonstrate that the relativistic model is still well-defined in the sense that the ground state energy is unique and bounded from below.  We consider finite-length closed curves. Neither are they intersecting, nor self-intersecting. The curves under consideration do not have any dynamical degrees of freedom. Moreover, for any given curve only one generic coupling constant is used  to define the strength of the interaction. Therefore, we will analyze how a relativistic Klein-Gordon particle interacts with singular interactions supported by stationary curves. This is the limit in which the interaction is strong enough (compared to the mass of the particle) that the relativistic effects can not be ignored but not strong enough to lead to pair production. The main motivation behind this work is to understand renormalization in depth, to see how well the effective interaction picture works in a simple model.
 
The construction of the renormalized model is based  on the renormalization of the principal operator. We will show that the renormalization of the model on a manifold can again be performed nonperturbatively as having been done in the nonrelativistic case in Ref.~\cite{kt}. Although the manifold in question being compact or noncompact plays a role while using upper bounds for the heat kernel, this distinction does not change our main results substantially. Therefore, whenever this contrast is needed in our calculations, this point will be emphasized. We also stress the fact that our method will cure the divergence in the codimension $2$ case in general. New divergences will appear in the case higher than codimension $2$, which can not be rendered finite through this prescription. We will see that one can apply distinct prescriptions in order to renormalize the principal operator for diverse aims. It will be observed that one needs to choose a certain prescription for the discussion of the renormalizability of the model, whereas it is more appropriate to utilize another prescription in order to study the renormalization group equation. If we think of renormalization as a way of writing down a sensible effective theory, the present study shows that this can be done successfully, that is keeping all the desired features of quantum mechanics intact. The details of the small scale physics are encoded, in some sense, by a minimal modification of the Hamiltonian.

The organization of the paper is as follows: in Section~\ref{cons}, we will construct the model following Ref.~\cite{kt}, and the renormalizability of the model will be studied. In Section~\ref{uniq}, the uniqueness of the ground state energy will be shown by using the Feynman-Hellman theorem at first. Second, that the ground state energy is bounded from below will be proven through the Ger\u{s}gorin theorem. In Section~\ref{rg}, the renormalization group equation will be obtained, and it will be demonstrated that the model is asymptotically free as its beta function is found to be negative.

\section{Construction of the renormalized model} \label{cons}
In this section, the construction of the renormalized model will be studied. We will consider a relativistic Klein-Gordon particle with a singular interaction, whose support is an arc length parametrized closed curves $\gamma(s)$ of length $L$ embedded in a $3$-dimensional Riemannian manifold. In general, we may have various non-intersecting curves, which we label as $\Gamma_i$, and their couplings may be  labeled as $\lambda_i$, as well. We assume that each $\Gamma_i$ has a finite length, and they are not self-intersecting. Moreover, they have to stay a certain minimum distance away from each other, which we may call $d_{ij}$, and they are only allowed to come close to themselves up to a certain distance, let us say $\Delta$, as we will make more precise later on. We will consider the following Hamiltonian as the cut-off regularized Hamiltonian of the model for the multi-curve case,
\begin{align}
H_\epsilon = \sum_\sigma \omega_\sigma a_\sigma^\dag a_\sigma - \sum_i \frac{\lambda_i}{L_i} \vert \Gamma_i^{\epsilon (-)} \rangle \langle \Gamma_i^{\epsilon(+)} \vert \,, 
\end{align}
where $\omega_\sigma^2 = f(\sigma) + m^2$, and 
\begin{align}
\vert \Gamma_i^{\epsilon (-)} \rangle = \sum_ \sigma \int_\mathcal{M} d_g^3 x \int_{\Gamma_i} d_g s K_{\epsilon/2} \left( \gamma(s), x \right) \frac{\phi_\sigma^*(x)}{\sqrt{2 \omega_\sigma}} a_\sigma^\dag \,.
\end{align}
$\langle \Gamma_i^{\epsilon (+)} \vert$ is similarly given in terms of the annihilation operator $a_\sigma$ instead of the creation operator $a_\sigma^\dag$. In the limit as $\epsilon^+ \rightarrow 0$, we get a delta function supported on the curve. Here, $\phi_\sigma(x)$ are the orthonormal complete (generalized) eigenfunctions of the Laplace-Beltrami operator $-\nabla_g^2$, and they satisfy the following equations,
\begin{align}
- \nabla_g^2 \phi_\sigma(x) &= f(\sigma) \phi_\sigma(x) \,,\\
\int_\mathcal{M} d_g^3 x \,\phi_\sigma^*(x)\phi_{\rho}(x) &= \delta_{\sigma \rho} \,,\\
\sum_{\sigma} \phi_\sigma^*(x) \phi_\sigma(y) &= \delta_g (x,y) \,. 
\end{align}
These equations are precise on a compact manifold. They should be interpreted in a generalized sense on a geodesically complete noncompact manifold since the Laplace-Beltrami operator is self-adjoint on these manifolds. Following Ref.~\cite{kt}, we need to obtain the regularized resolvent so that the principal operator can be used to renormalize the theory. The regularized resolvent is given by
\begin{align}
(H - E)^{-1} &= (H_0 - E)^{-1} + \frac{1}{\sqrt{L_i L_j}} (H_0 - E)^{-1} \vert \Gamma_i^{\epsilon(-)} \rangle \Phi_{ij}^{-1} \langle \Gamma_j^{\epsilon(+)} \vert (H_0 - E)^{-1} \,,  
\end{align} 
where a summation is assumed over the repeated indices, and $\Phi$ refers  to the principal operator. It is compactly written as
\begin{align}
\Phi_{ij} &= \left\{ \begin{array}{l} \frac{1}{\lambda_i} - \frac{1}{L_i} \langle \Gamma_i^{\epsilon(+)} \vert (H_0 - E)^{-1} \vert \Gamma_i^{\epsilon(-)} \rangle \,, \\ - \frac{1}{\sqrt{L_i L_j}} \langle \Gamma_i^{\epsilon(+)} \vert (H_0 - E)^{-1} \vert \Gamma_j^{\epsilon(-)} \rangle \,. \end{array} \right.
\end{align}
It is obvious that the off-diagonal part of the principal operator does not need any renormalization. However, we can not say the same thing for the on-diagonal part of it. We will give the explicit form of the latter for the single curve case just for simplicity, and its generalization to multi-curve is obvious. The regularized principal operator in that case is, then, given by
\begin{align}
\Phi_\epsilon (E) &=  \frac{1}{\lambda(\epsilon)} - \frac{1}{L} \iint_\mathcal{M} d_g^3 x \, d_g^3 y \iint_{\Gamma \times \Gamma} d_g s \, d_g s'  K_{\epsilon/2} \left( \gamma(s) ,x \right) K_{\epsilon/2} \left( y , \gamma(s') \right) \nonumber \\
& \quad \times \sum_{\rho,\sigma}\frac{\phi_\rho(x)\phi_\sigma^*(y)}{\sqrt{4 \omega_\rho \omega_\sigma}} a_\rho \frac{1}{H_0 - E} a_\sigma^\dag \,.
\end{align}
If we put this expression into normal ordered form by moving the creation operator to the left of all the annihilation operators and vice versa, this expression is modified, and takes the following form,
\begin{align}
\Phi_\epsilon (E) &= \frac{1}{\lambda(\epsilon)} - \frac{1}{L} \iint_\mathcal{M} d_g^3 x \, d_g^3 y \iint_{\Gamma \times \Gamma} d_g s \, d_g s'  K_{\epsilon/2} \left( \gamma(s) ,x \right) K_{\epsilon/2} \left( y , \gamma(s') \right) \nonumber \\
& \quad \times \sum_{\rho,\sigma}\frac{\phi_\rho(x)\phi_\sigma^*(y)}{\sqrt{4 \omega_\rho \omega_\sigma}} \left[ \frac{\delta_{\rho \sigma}}{H_0 - E + \omega_\sigma} + a_\sigma^\dag \frac{1}{H_0 - E + \omega_\rho + \omega_\sigma} a_\rho \right] \,.
\end{align}
Since the ground state corresponds to a single boson state, the principal operator becomes 
\begin{align}
\Phi_\epsilon (E) &= \frac{1}{\lambda(\epsilon)} - \frac{1}{2L} \iint_\mathcal{M} d_g^3 x \, d_g^3 y \iint_{\Gamma \times \Gamma} d_g s \, d_g s'  K_{\epsilon/2} \left( \gamma(s) ,x \right) K_{\epsilon/2} \left( y , \gamma(s') \right) \nonumber \\
& \quad \times \sum_\sigma\frac{\phi_\sigma(x)\phi_\sigma^*(y)}{ \omega_\sigma} \frac{1}{- E + \omega_\sigma} \,.
\end{align}
Now, we need to take care of the fractions. After a Feynman parametrization and exponentiation, we can place $\omega_\sigma$ onto the exponent. If one uses the subordination identity below,
\begin{align}
e^{- \omega_\sigma t} = \frac{t}{2 \sqrt{\pi}} \int_0^\infty d u \, \frac{e^{- t^2 / 4 u - \omega_\sigma^2 u}}{u^{3/2}} \,,
\end{align}
it is, then, easy to recognize the following eigenfunction expansion of the heat kernel,
\begin{align}
K_u (x,y) &= \sum_\sigma e^{- f(\sigma) u} \phi_\sigma^*(x) \phi_\sigma(y) \,.
\end{align}
After applying the subordination identity, doing a partial integration, recognizing the eigenfunction expansion of the heat kernel, utilizing the semigroup property of the heat kernel, and shifting the variable $u$ successively, the following expression is obtained,
\begin{align} \label{phi}
\Phi_\epsilon (E) &= \frac{1}{\lambda(\epsilon)} - \frac{1}{2\sqrt{\pi}L} \iint_{\Gamma \times \Gamma} d_g s \, d_g s' \int_0^\infty dt \, e^{E t} \int_\epsilon^\infty du \frac{e^{- m^2(u-\epsilon) - t^2/4 (u-\epsilon)}}{\sqrt{u-\epsilon}} K_u \left( \gamma(s) , \gamma(s') \right) \,.  
\end{align}
Let us choose the coupling constant in such a way that after taking the limit $\epsilon \rightarrow 0^+$, the divergence can be eliminated,
\begin{align}
\frac{1}{\lambda(\epsilon)} = \frac{1}{\lambda_R(\mu)} + \frac{1}{2\sqrt{\pi}L} \iint_{\Gamma \times \Gamma} d_g s \, d_g s' \int_0^\infty dt \, e^{\mu t} \int_\epsilon^\infty du \frac{e^{- m^2(u-\epsilon) - t^2/4 (u-\epsilon)}}{\sqrt{u-\epsilon}} K_u \left( \gamma(s) , \gamma(s') \right) \,,
\end{align}
where $\lambda_R(\mu)$ is the renormalized coupling constant, and $\mu$ is the renormalization scale. After plugging this redefinition into Eq.~(\ref{phi}) and rescaling the variable $t \rightarrow \sqrt{u} t$, the renormalized principal operator is, then, given by
\begin{align} \label{rp}
\Phi_R (E) &= \frac{1}{\lambda(\mu)} + \frac{1}{2\sqrt{\pi}L} \iint_{\Gamma \times \Gamma} d_g s \, d_g s' \int_0^\infty dt \, e^{- t^2/4} \nonumber \\
& \quad \times \int_0^\infty du e^{-m^2 u} \left( e^{\mu t \sqrt{u}} - e^{E t \sqrt{u}} \right) K_u \left( \gamma(s) , \gamma(s') \right) \,.
\end{align}
Here, the renormalization scale $\mu$ is chosen to be greater than the energy $E$, and the energy $E$ is restricted to the following range,
\begin{align}
-m < E < m\,.
\end{align}
This renormalization scale could be eliminated in favor of a physical parameter via imposing a renormalization condition, e.g. a physical bound state energy. In that case, the renormalization scale should be chosen to be equal to the physical bound state $E_b$ in order to obtain the zero eigenvalues of the renormalized principal operator due to the fact that the zeros of this operator correspond to the poles of the resolvent operator. Therefore, we can directly read the physical bound state energy from this equation as
\begin{align}
\Phi_R(E_b) = 0 \,.
\end{align}    

From this point on, we will utilize a minimal prescription, in which $\mu$ will be set equal to the binding energy of a given curve and the inverse of renormalized coupling will be sent to zero. This is the most convenient choice from the point of view of bound state problems. 

As in Ref.~\cite{kt}, in order to prove the renormalizability of the model, we should also observe that the expression below stays finite in the limit $s \rightarrow s'$ due to the occurrence of a possible singularity of the heat kernel in this limit, 
\begin{align}
\frac{1}{2\sqrt{\pi}L} \iint_{\Gamma \times \Gamma} d_g s \, d_g s' \int_0^\infty dt \, e^{- t^2/4} \int_0^\infty du \, e^{-m^2 u} \left( e^{\mu t \sqrt{u}} - e^{E t \sqrt{u}} \right) K_u \left( \gamma(s) , \gamma(s') \right) < \infty \, .
\end{align}
In that respect, we will push the value of our expression all the way to much  larger values until we find an upper bound of the expression, which should, obviously, be large but finite. Henceforward, we assume that our manifold $\mathcal{M}$ belongs to a class which admits certain types of upper and lower bounds on  heat kernels as discussed in Chapters~$15$ and~$16$ in Ref.~\cite{gri}. For noncompact manifolds, this expression can be made larger by replacing the heat kernel by a bigger quantity via an off-diagonal upper bound for the heat kernel as stated in Section~$4$ in Ref.~\cite{et}, which is based on the aforementioned chapters in Ref.~\cite{gri}. One, then, obtains 
\begin{align}
\frac{B}{16 \pi^2 L} \iint_{\Gamma \times \Gamma} d_g s \, d_g s' \int_0^\infty dt \, e^{- t^2/4} \int_0^\infty du \, e^{-m^2 u} \left( e^{\mu t \sqrt{u}} - e^{E t \sqrt{u}} \right) \frac{e^{-d_g^2\left( \gamma(s),\gamma(s') \right)/4 C u}}{u^{3/2}} \,,
\end{align}
in which $d_g$ is the geodesic distance, $B$ and $C$ are some constants, whose values specifically depend on the geometry of the manifolds in question, but not important for the sake of the renormalizability. For compact manifolds, there is a term proportional to $1/V(\mathcal{M})$ in the off-diagonal upper bounds as stated in the aforesaid section in Ref.~\cite{et}, which is based on Ref.~\cite{wa} and the chapters mentioned above in Ref.~\cite{gri}. However, it gives a finite contribution. Taking the $t$-integral gives
\begin{align} \label{dg}
& \frac{B}{16 \pi^2 L} \iint_{\Gamma \times \Gamma} d_g s \, d_g s' \int_0^\infty du  \frac{e^{-d_g^2\left( \gamma(s),\gamma(s') \right)/4 C u}}{u^{3/2}} \nonumber \\
& \qquad \quad \times \left\{ e^{-(m^2 - \mu^2) u} \left[\sqrt{\pi} + 2 \mu \int_0^1 dt \, e^{-\mu^2 u t^2}\right] - e^{-(m^2 - E^2) u} \left[\sqrt{\pi} + 2 E \int_0^1 dt \, e^{-E^2 u t^2}\right] \right\} \,.
\end{align}
Since the singularity occurs $s \rightarrow s'$, we would like to rewrite this expression by dividing it into the part which can generate this singularity, and the rest. Showing that the former stays finite implies that the expression around the whole curve is free of divergences. Eq.~(\ref{dg}) can be rewritten as
\begin{align} \label{sum}
& \frac{B}{16 \pi^2 L} \int_\Gamma d_g s \left( \int_{\vert \xi \vert < \delta} d_g \xi + \int_{\vert \xi \vert > \delta} d_g \xi \right) \int_0^\infty du  \frac{e^{-d_g^2\left( \gamma(s),\gamma(s') \right)/4 C u}}{u^{3/2}} \nonumber \\
& \qquad \quad \times \left\{ e^{-(m^2 - \mu^2) u} \left[\sqrt{\pi} + 2 \mu \int_0^1 dt \, e^{-\mu^2 u t^2}\right] - e^{-(m^2 - E^2) u} \left[\sqrt{\pi} + 2 E \int_0^1 dt \, e^{-E^2 u t^2}\right] \right\} \,,
\end{align}
in which $\xi$ is the arc length of the curve between the points $s$ and $s'$, defined as $\xi = \vert s-s' \vert$ for $\xi \in [- L/2, L/2]$. $\delta$ is the maximum value, which $\xi$ can get in a geodesic ball centered at the point $s$. In Eq.~(\ref{sum}), the part between the points $[-\delta,\delta]$ is actually responsible for the possible divergence, so we will focus on that part, henceforth. In Ref.~\cite{kt}, the authors obtained the following inequality, which relates the arc length of a segment of the curve to its curvature, 
\begin{align}
\sqrt{1-\kappa_g^* \delta} \xi &< d_{g} \left( \gamma(s) ,\gamma(s') \right) < \xi \text{ with }0 < \delta < 1 / 2 \kappa_g^* \,. 
\end{align}
We can make the part which is responsible for the possible divergence larger through this inequality,
\begin{align} \label{div}
& \frac{B}{8 \pi^2 L} \int_\Gamma d_g s \int_0^\delta d_g \xi \int_0^\infty du  \frac{e^{-(1- \kappa_g^* \delta)\xi^2/4 C u - m^2 u}}{u^{3/2}} \nonumber \\
& \qquad \quad \times \left\{ e^{-(m^2 - \mu^2) u} \left[\sqrt{\pi} + 2 \mu \int_0^1 dt \, e^{-\mu^2 u t^2}\right] - e^{-(m^2 - E^2) u} \left[\sqrt{\pi} + 2 E \int_0^1 dt \, e^{-E^2 u t^2}\right] \right\} \,.
\end{align}
On the other hand, another condition is needed to be imposed on the curve, which prevents the curve to get close to itself immediately after leaving the neighborhood around the center of the geodesic ball. This can be stated as
\begin{align}
d_g \left( \gamma(s),\gamma(s') \right) > \Delta \text{ if } \xi > \delta \,.
\end{align}
Here, $\Delta$ can be smaller than $\delta$. Extending the upper limit of the $\xi$-integral to $\infty$ even makes the expression larger. After straightforward calculations, the terms without the $t$-integrals in Eq.~(\ref{div}) take the following form
\begin{align} \label{not}
\frac{B \sqrt{C}}{4 \pi \sqrt{1- \kappa_g^* \delta}} \log \sqrt{\frac{m^2 - E^2}{m^2 - \mu^2}} \,,
\end{align}
for which either one of the following conditions $(\mu<0$ $\wedge$ $m+\mu>0)$ $\vee$ $(\mu>0$ $\wedge$ $m>\mu)$ should be satisfied, and the same for $E$, as well.

On the other hand, by taking the $u$-integral at first for the rest in Eq.~(\ref{div}), the following expression is obtained,
\begin{align}
& \frac{B}{2 \pi^2 L} \int_\Gamma d_g s \int_0^\infty d_g \xi \int_0^1 dt \left[ \mu K_0 \left(\sqrt{\frac{1-\kappa_g^* \delta}{C}} \sqrt{m^2-(1-t^2)\mu^2} \xi\right) \right. \nonumber \\
& \left. \qquad - E K_0 \left(\sqrt{\frac{1-\kappa_g^* \delta}{C}} \sqrt{m^2-(1-t^2)E^2} \xi \right)\right] \,,
\end{align}
where $K_0(x)$ is the modified Bessel function of the second kind of order zero~\cite{ol}. After taking the integrals with the conditions $-m<\mu<m$ $\wedge$ $-m<E<m$, and combining them with the result~(\ref{not}), the whole expression takes the following form,
\begin{align}
\frac{B \sqrt{C}}{4 \pi \sqrt{1- \kappa_g^* \delta}} \left[ \log \sqrt{\frac{m^2 - E^2}{m^2 - \mu^2}}
+ \mathrm{sgn}(\mu) \log \left( \frac{m+|\mu|}{\sqrt{m^2-\mu^2}} \right) - \mathrm{sgn}(E) \log \left( \frac{m+|E|}{\sqrt{m^2-E^2}} \right) \right] \,,
\end{align}
which is  positive and finite, hence it implies that the model is renormalizable.
\section{Uniqueness of the ground state energy and existence of a lower bound for it} \label{uniq}
In order to show that the ground state energy is unique, we will resort to the well-known Feynman-Hellman theorem~\cite{fh} so as to see how the eigenvalues of the principal operator flow with respect to the change in the energy variable $E$. Let us assume that $\Phi_{ij}$ satisfies the following eigenvalue equation for the $k$th eigenvalue $\omega^{(k)}(E)$,
\begin{align}
\Phi_{ij}(E) A_j^{(k)} &= \omega^{(k)}(E) A_i^{(k)} \,,
\end{align}
$A^{(k)}$ being the $k$th eigenvector, and the summation convention over the repeated index $j$ is adopted. The derivative of the principal operator in Eq.~(\ref{rp}) with respect to the energy $E$ reads
\begin{align}
\frac{\partial \Phi_{ij} (E)}{\partial E} &= - \frac{1}{2\sqrt{\pi L_i L_j}} \iint_{\Gamma_i \times \Gamma_j} d_g s \, d_g s' \int_0^\infty dt \, t e^{- t^2/4} \nonumber \\
& \qquad \times \int_0^\infty du \, \sqrt{u} e^{-m^2 u + E t \sqrt{u}} K_u \left( \gamma_i(s) , \gamma_j(s') \right) \,.
\end{align}
By the Feynman-Hellman theorem, the expectation value of this expression should be equal to the derivative of the eigenvalues of the operator with respect to the energy $E$, which we rewrite in a convenient form,
\begin{align}
\frac{\partial \omega^{(k)} (E)}{\partial E} &= - \frac{1}{2\sqrt{\pi}} \int_\mathcal{M} d_g^3 x \int_0^\infty dt \, t e^{- t^2/4} \int_0^\infty du \, \sqrt{u} e^{-m^2 u + E t \sqrt{u}} \nonumber \\
& \qquad \times \left \vert \sum_i \frac{1}{\sqrt{L_i}}\int_{\Gamma_i} d_g s \, K_u \left( \gamma_i(s) , x \right) A_i^{(k)}\right \vert^2 \,.
\end{align}
The expression above is obviously strictly negative, hence all eigenvalues are decreasing functions of energy. This tells that the ground state energy must correspond to the zero of the lowest eigenvalue of the principal operator. Note that the principal operator has the following property (for our choice of renormalization prescription), all the diagonal elements are positive and all the off diagonal terms are strictly negative and it is a Hermitian matrix. In such a case, by applying the Trotter product formula we can see that it is an ergodic matrix, thus has a unique lowest eigenvalue, whose eigenvector can be chosen with all positive components. Hence, this implies the uniqueness of the ground state energy.

As in Ref.~\cite{kt}, we will use the Ger\u{s}gorin theorem~\cite{mat} in order to show that there exists a lower bound for the ground state energy. This theorem states that all the eigenvalues $\omega$ of a matrix $\Phi \in M_N$ are located in the union of $N$ disks:
\begin{align}
\bigcup_{i=1}^N \vert \omega - \Phi_{ii} \vert \leqslant \bigcup_{i=1}^N \sum_{i \neq j = 1}^N \vert \Phi_{ij} \vert \,.
\end{align}
If there is a lower bound on the ground state energy, say $E^*$ in our problem, one expects that $\omega = 0$ is not the eigenvalue at all beyond this lower bound, $E \leqslant E^*$. In Ref.~\cite{et}, it is shown that the following inequality
\begin{align} \label{inge}
\vert \Phi_{ii} (E) \vert^{\min} > (N-1) \vert \Phi_{ij} (E) \vert^{\max}_{i\neq j} \,, 
\end{align}
implies that the ground state energy should be larger than a lower bound $E^*$, which saturates the above inequality,
\begin{align}
E_{gr} \geqslant E^* \,.
\end{align}
So as to obtain a lower bound for the on-diagonal part of the principal operator for the left-hand side of the inequality~(\ref{inge}), one needs to use an off-diagonal lower bound for the heat kernel in Chapter $5.6$ in Ref.~\cite{dav} or the one in Ref.~\cite{cou} when certain conditions imposed on the manifold are satisfied. Therefore, we do not here specify the type of the manifold for the calculation of the lower bound for the on-diagonal part of the principal operator. The length of a geodesic is known to be less than or equal to any admissible curve with the same endpoints. Thereof, replacing the heat kernel by a smaller quantity via the lower bound, and then replacing the geodesic distance $d_g$ by the arc length $\xi$, the diagonal part of the principal operator becomes smaller as
\begin{align} \label{pii}
\left\vert \Phi_{ii}(E) \right\vert > \frac{A}{8 \pi^2 L_i} \int_{\Gamma_i} d_g s \int_0^{L_i/2} d_g \xi \int_0^\infty d t \, \left( e^{\mu_i t} - e^{E t} \right) \int_0^\infty d u \frac{e^{-m^2 u - t^2/4 u - \xi^2/ 4 D u}}{u^2} \,,
\end{align}
where $A$ and $D$ are some constants. Taking the $u$-integral first with some rescaling gives
\begin{align} \label{pii2}
\frac{A m}{8 \pi^2 L_i} \int_\Gamma d_g s \int_0^{L_i/2} d_g \xi \int_0^\infty d t \, \frac{ e^{\mu_i t \xi /\sqrt{D}} - e^{E t \xi /\sqrt{D}}}{\sqrt{1+t^2}} K_1 \left( \frac{m \xi}{\sqrt{D}} \sqrt{1+t^2} \right)  \,.
\end{align}
Dropping $1$ in the integral representation of the modified Bessel function of the second kind in Ref.~\cite{gr},
\begin{align} \label{k1}
K_\nu(x z) = \sqrt{\frac{\pi}{2 z}} \frac{x^\nu e^{-x z}}{\Gamma\left(\nu + \frac{1}{2}\right)} \int_0^\infty d t \, e^{-x t} t^{\nu - 1/2} \left( 1 + \frac{t}{2 z} \right)^{\nu - 1/2} \,, \quad | \mathrm{arg} z| < \pi  \,, \Re \nu > - \frac{1}{2} \,, x > 0    \,,
\end{align}
leads to the following lower bound for $K_1(z)$,
\begin{align}
K_1(z) > \frac{e^{-z}}{z} \,.
\end{align}
After using this lower bound in the expression~(\ref{pii2}), the denominator becomes $1+t^2$. Replacing $1+t^2$ by $(1+t)^2$ even makes the expression smaller, so we obtain
\begin{align}
\frac{A \sqrt{D}}{8 \pi^2 L_i} \int_\Gamma d_g s \int_0^\infty \frac{dt}{(1+t)^2}\int_0^{L_i/2} d_g \xi \, e^{- m \xi / \sqrt{D}} \frac{ e^{-(m-\mu_i) t \xi /\sqrt{D}} - e^{-(m-E) t \xi /\sqrt{D}}}{\xi} \,.
\end{align}
Now, we rewrite the integral over $\xi$ as $\int_0^\infty - \int_{L_i/2}^\infty$. Taking the integrals in the same order in the  the integral $\int_0^\infty$ leads to
\begin{align} \label{i1}
\frac{A \sqrt{D}}{8 \pi^2} \left[ \frac{m \log m - (m-E) \log(m-E)}{E} - \frac{m \log m - (m-\mu_i) \log(m-\mu_i)}{\mu_i} \right] \,.
\end{align}
In the second integral, we will replace $\xi$ by $L_i/2$ in the exponent of the exponential term which does not have $t$ inside. Afterward, we will extend to the lower limit to $\infty$ in order to make this integral larger so that the expression~(\ref{pii2}) becomes even more smaller. We obtain the following result for the second integral after calculating  all the integrals,
\begin{align} \label{i2}
\frac{A \sqrt{D}}{16 \pi} e^{-m L_i/2 \sqrt{D}} \log \frac{m-E}{m-\mu_i} \,.
\end{align}
By combining the results~(\ref{i1}) and~(\ref{i2}), we have the lower bound estimate for the absolute value of the on-diagonal part of the principal operator,
\begin{align}
|\Phi_{ii} (E)| > \frac{A \sqrt{D}}{8 \pi^2} \left[ \left( 1 - \frac{m}{\mu_i}- \frac{\pi}{2} e^{-m L_i / 2\sqrt{D}} \right) \log \frac{m-E}{m-\mu_i} + \left( \frac{m}{E}-\frac{m}{\mu_i} \right) \log \frac{m}{m-E} \right] \,.
\end{align}
By introducing $L_{\min} =\min_i (L_i)$ and $\mu_{\min} = \min_i (\mu_i)$, and rearranging the above inequality, the lower bound of the right-hand side can be given uniformly as follows, 
\begin{align}
|\Phi_{ii} (E)| > \frac{A \sqrt{D}}{8 \pi^2} \left[ \left( 1 - \frac{\pi}{2} e^{-m L_{\min}/2 \sqrt{D}} \right) \log \frac{m-E}{m-\mu_{\min}} + \frac{m}{E} \log \frac{m}{m-E} - \frac{m}{\mu_{\min}} \log \frac{m}{m-\mu_{\min}} \right] \,.
\end{align}
Here we search for the solution $E^*$ in the range $(-m, \mu_{min})$, therefore for sufficiently large values of $mL_{\min}$ and for $\mu_{\min}$ not too close to $-m$ the above expression is always positive.
 
We also need to have an upper bound on the absolute value of the off-diagonal part of the principal operator for the inequality~(\ref{inge}). Since the behavior of the off-diagonal upper bounds on heat kernels for compact manifolds differ from the ones for noncompact manifolds with a term proportional to the volume of the manifold, we will estimate the bound for compact manifolds. For the noncompact case, the volume term automatically drops. It can easily be shown that the absolute value of that part is given by
\begin{align}
\left\vert \Phi_{ij}(E) \right\vert = \frac{1}{2 \sqrt{\pi} \sqrt{L_i L_j}} \iint_{\Gamma_i \times \Gamma_j} d_g s \, d_g s' \int_0^\infty d t \, e^{E t} \int_0^\infty d u \frac{e^{-m^2 u - t^2/4 u}}{\sqrt{u}} K_u\left( \gamma_i(s), \gamma_j(s') \right) \,.
\end{align}
By means of the off-diagonal upper bound for the heat kernel in Ref.~\cite{gri}, which we similarly utilized in the previous section, we can enlarge the expression above,
\begin{align}
\left\vert \Phi_{ij}(E) \right\vert &< \frac{F}{2 \sqrt{\pi} \sqrt{L_i L_j} V(\mathcal{M})} \iint_{\Gamma_i \times \Gamma_j} d_g s \, d_g s' \int_0^\infty d t \, e^{E t} \int_0^\infty d u \frac{e^{-m^2 u - t^2/4 u - d_g^2 \left( \gamma_i(s), \gamma_j(s') \right)/4 C u}}{\sqrt{u}} \nonumber \\
& \quad + \frac{B}{16 \pi^2 \sqrt{L_i L_j}} \iint_{\Gamma_i \times \Gamma_j} d_g s \, d_g s' \int_0^\infty d t \, e^{E t} \int_0^\infty d u \frac{e^{-m^2 u - t^2/4 u - d_g^2 \left( \gamma_i(s), \gamma_j(s') \right)/4 C u}}{u^2} \,.
\end{align}
Introducing the following minimum distance between the curves allows us to make the right-hand side larger.
\begin{align}
d_{ij} &= d ( \Gamma_i, \Gamma_j) = \min_{s \in \Gamma_i,s' \in \Gamma_j} d_g ( \gamma_i(s), \gamma_j(s') ) \,.
\end{align}
After taking the $s$- and $u$-integrals with the replacement of $d_g$ by $d_{ij}$, we obtain 
\begin{align} \label{pij}
\left\vert \Phi_{ij}(E) \right\vert &< \frac{F d_{ij} \sqrt{L_i L_j}}{2 V(\mathcal{M}) \sqrt{C} m} \int_0^\infty d t \, e^{E d_{ij} t /\sqrt{C} - m d_{ij} \sqrt{1+t^2} / \sqrt{C}} \nonumber \\
& \quad + \frac{B m \sqrt{L_i L_j}}{4 \pi^2} \int_0^\infty d t \, \frac{e^{E d_{ij} t /\sqrt{C}}}{\sqrt{1+t^2}} K_1 \left( \frac{m d_{ij}}{\sqrt{C}} \sqrt{1+t^2} \right) \,.
\end{align}
Via the same integral representation of the modified Bessel function $K_1(z)$~(\ref{k1}), we can convert the second line in the above expression into
\begin{align}
\frac{B C^{1/4} m \sqrt{L_i L_j}}{2 \sqrt{2} \pi^2 \sqrt{m d_{ij}}} \int_0^\infty d t \, \frac{e^{E d_{ij} t /\sqrt{C} - m d_{ij} \sqrt{1+t^2} / \sqrt{C}}}{(1+t^2)^{3/4}} \int_0^\infty d u \, e^{-u} \sqrt{u} \left( 1 + \frac{\sqrt{C} u}{2 m d_{ij} \sqrt{1+t^2}} \right)^{1/2} \,.
\end{align}
The full expression~(\ref{pii}) is smaller than the following one,
\begin{align}
& \frac{F d_{ij} \sqrt{L_i L_j}}{2 V(\mathcal{M}) \sqrt{C} m} \int_0^\infty d t \, e^{-(m-E) d_{ij} t /\sqrt{C}} \nonumber \\
& \quad + \frac{B C^{1/4} m \sqrt{L_i L_j}}{2 \sqrt{2} \pi^2 \sqrt{m d_{ij}}} \int_0^\infty d t \, e^{-(m-E) d_{ij} t / \sqrt{C}} \int_0^\infty d u \, e^{-u} \sqrt{u} \left( 1 + \frac{\sqrt{C} u}{2 m d_{ij}} \right)^{1/2} \,. 
\end{align}
Calculating the integrals gives
\begin{align}
\left\vert \Phi_{ij}(E) \right\vert < \frac{F}{2} \frac{1}{m(m-E)} \frac{\sqrt{L_i L_j}}{V(\mathcal{M})} + \frac{B \sqrt{C}}{4 \pi^2} \frac{m}{m-E} \frac{\sqrt{L_i L_j}}{d_{ij}} e^{m d_{ij} / \sqrt{C}} K_1 \left( \frac{m d_{ij}}{\sqrt{C}} \right) \,.
\end{align}
If one uses another integral representation of the modified Bessel function of the second kind in Ref.~\cite{ol},
\begin{align}
K_\nu (z) = \frac{\sqrt{\pi}(\frac{z}{2})^\nu}{\Gamma \left( \nu + \frac{1}{2} \right)} \int_1^\infty d t \, e^{-z t} \left( t^2 - 1 \right)^{\nu - 1/2} \,, \quad \Re \nu> - \frac{1}{2} \,, |\mathrm{arg} z| < \frac{\pi}{2} \,, 
\end{align}
then the following upper bound is obtained for $K_1$ when we replace $t^2-1$ by $t^2$,
\begin{align}
K_1 (z) < e^{-z} \left( 1 + \frac{1}{z} \right) \,.
\end{align}
Through this inequality, the off-diagonal upper bound for the absolute value of the principal operator takes the following form
\begin{align}
\left\vert \Phi_{ij}(E) \right\vert < \frac{F}{2} \frac{1}{m(m-E)} \frac{\sqrt{L_i L_j}}{V(\mathcal{M})} + \frac{B \sqrt{C}}{4 \pi^2} \frac{m}{m-E} \frac{\sqrt{L_i L_j}}{d_{ij}} \left( 1 + \frac{\sqrt{C}}{m d_{ij}} \right) \,.
\end{align}
By introducing $d_{\min} = \min_{i \neq j} (d_{ij})$ and $L_{\max} =\max_i (L_i)$, we maximize the right-hand side of the above inequality,
\begin{align}
\left\vert \Phi_{ij}(E) \right\vert^{\max} < \frac{F}{2} \frac{1}{m(m-E)} \frac{L_{\max}}{V(\mathcal{M})} + \frac{B \sqrt{C}}{4 \pi^2} \frac{m}{m-E} \frac{L_{\max}}{d_{\min}} \left( 1 + \frac{\sqrt{C}}{m d_{\min}} \right) \,.
\end{align}
As a result, the inequality~(\ref{inge}) becomes
\begin{align}
& \left( 1 - \frac{\pi}{2} e^{-m L_{\min}/2 \sqrt{D}} \right) \log \frac{m-E}{m-\mu_{\min}} + \frac{m}{E} \log \frac{m}{m-E} - \frac{m}{\mu_{\min}} \log \frac{m}{m-\mu_{\min}}
\nonumber \\
& \qquad > (N-1)\left[ \frac{4 \pi^2 F}{A \sqrt{D}} \frac{1}{m(m-E)} \frac{L_{\max}}{V(\mathcal{M})} + 2 \sqrt{\frac{B^2 C}{A^2 D}} \frac{m}{m-E} \frac{L_{\max}}{d_{\min}} \left( 1 + \frac{\sqrt{C}}{m d_{\min}} \right) \right] \,.
\end{align}
As explained before, the term proportional to the volume of the manifold automatically drops for noncompact manifolds. As one can see by inspection,  for large separations (compared to $m$) and sufficiently long (compared to $m$) curves, we have solutions $-m<E<E^*$ satisfying the above inequality.
\section{Renormalization group equation} \label{rg}
In this section, we will derive the beta function of the model, and show that the principal operator satisfies the renormalization group equation. Another renormalization prescription that is different from the one which we use to show that the model is renormalizable in Section~\ref{cons} will be chosen so as to achieve the goals of this section as it was done in Ref.~\cite{kt}. As our model consists of closed curves embedded in Riemannian manifolds, the curves can be treated as various embedded circles, and one can locally choose orthonormal frame bundles over the Riemannian manifold, which carry adapted frames of the curves~\cite{wil}. Those adapted frames are decomposed into a vector tangent to the curves, and the rest normal to the curves. Since the divergence only occurs while moving along the curve in the coincidence limit $s \rightarrow s'$, there is no geometric contribution to this divergence, coming from the directions related to the normal bundle. This means that the heat kernel can at least locally be written as a direct product of heat kernels as long as the coincidence limit is concerned. These heat kernels are nothing but the one on a circle with the induced metric, coming from the embedding, and the one coming from the normal directions. That we are only moving along the curve allows us to choose locally the heat kernel of $\mathbb{R}^2$ in the normal directions. For the removal of the singularity in the expression, we will use the heat kernel on a circle with the induced metric, coming from the embedding. Henceforward, we are working with a single curve for simplicity, however, the generalization to the multi-curve case is obvious. In the light of the above discussion, the following form is chosen so as to renormalize the principal operator for the single curve case,
\begin{align}
\Phi_R(E) &= \frac{1}{\lambda_R(\mu)} + \frac{1}{2 \sqrt{\pi}} \frac{1}{\int_\Gamma d_g s} \int_0^\infty d t \int_0^\infty d u \, e^{-t^2 / 4 u} \nonumber \\
& \quad \times \left[ \iint_{\Gamma \times \Gamma} d_g \theta \, d_g \theta' \frac{e^{- \mu t}}{4 \pi u^{3/2}} K_u^{S^1} \left( \gamma(\theta), \gamma(\theta'); e^*(g) \vert_{S^1} \right) \right. \nonumber \\
& \left. \qquad - \iint_{\Gamma \times \Gamma} d_g s \, d_g s' \frac{e^{E t - m^2 u}}{\sqrt{u}} K_u \left( \gamma(s),\gamma(s'); g \right) \right] \,,
\end{align}
in which $d_g \theta = \sqrt{e^*(g) \vert_{S^1}} d \theta$, and $e^*(g) \vert_{S^1}$ is a pull-back metric. The renormalization group equation allows us to obtain the beta function of the coupling constant,
\begin{align}
\mu \frac{d}{d \mu} \Phi_R (E)=0 \,.
\end{align}
If we plug the principal operator above into this equation, do the following rescaling $t \rightarrow \mu^{-1} t$, $u \rightarrow \mu^{-2} u$, and apply the rescaling property of the heat kernel, i.e. $K_u(x,y;g) = \mu^{-d} K_{\mu^{-2} u}(x,y;\mu^{-2} g)$, where $d$ is the dimension of the space, the beta function is obtained as
\begin{align} \label{be}
\beta (\lambda_R) &= -\frac{\lambda_R^2}{8 \pi^{3/2}} \frac{1}{\int_\Gamma d_{\mu^2 g} s} \iint_{\Gamma \times \Gamma} d_{\mu^2 g} \theta \, d_{\mu^2 g} \theta' \int_0^\infty d t \, t e^{-t} \nonumber \\
& \quad \times \int_0^\infty d u  \, \frac{e^{-t^2/4 u}}{u^{3/2}} K_u^{S^1} \left( \gamma(\theta), \gamma(\theta'); \mu^2 e^*(g) \vert_{S^1} \right) \,.
\end{align}
That the metric is invariant under the full group of Diff$(S^1)$ allows us to redefine the coordinate $\theta$ as
\begin{align}
\tilde{\theta} &= \int_0^\theta d \theta' \sqrt{e^*(g(\theta')) \vert_{S^1}} \,.
\end{align}
This redefinition removes the metric dependence in such a way that the metric can be converted into the standard metric on $S^1$, so the heat kernel in Eq.~(\ref{be}) is given by the standard heat kernel on $S^1$~\cite{ros} as 
\begin{align}
K_u ( \tilde{\theta},\tilde{\theta}') &= \sum_{n=-\infty}^\infty \frac{e^{-\frac{n^2 \pi^2 u}{L^2}} e^{\frac{i n \pi (\tilde{\theta} - \tilde{\theta}')}{L}}}{L} \,.
\end{align}
Thus, the beta function becomes metric independent, and only the length of the curve plays a distinctive role for the beta function. Thereof, the beta function is given by
\begin{align}
\beta (\lambda_R) = -\frac{\lambda_R^2}{8 \pi^{3/2}} \frac{1}{L} \iint_{\Gamma \times \Gamma} d \tilde{\theta} \, d \tilde{\theta}' \int_0^\infty d t \, t e^{-t} \int_0^\infty d u  \, \frac{e^{-t^2/4 u}}{u^{3/2}} K_u ( \gamma(\tilde{\theta}), \gamma(\tilde{\theta}') ) \,.
\end{align}
It is obvious that the model is asymptotically free as the integrals give a positive contribution. How the change of the energy scale effects the coupling constant can be observed by integrating the beta function. The flow of the coupling constant, hence, takes the following form,
\begin{align}
\lambda(\tau \mu) = \frac{\lambda(\mu)}{1+ \frac{\lambda(\mu)}{8 \pi^{3/2} L} C \log \tau} \,,
\end{align}
$C$ being a constant, and it stands for the result coming from the integrations.

In the following, we will demonstrate that the renormalized principal operator satisfies the renormalization group equation. In that purpose, it will be shown that the change of the renormalization scale is equivalent to the scaling of the energy, the mass, and the metric in the expression. In order to observe this, we make the following rescaling for the physical parameters in the principal operator,
\begin{align}
E \rightarrow \tau E \,, \quad  m \rightarrow \tau m\,, \quad g \rightarrow \tau^{-2} g \,.
\end{align}
Straightforward calculations give
\begin{align}
\Phi_R(\mu,\lambda_R(\mu),\tau E,\tau m,\tau^{-2} g) &= \frac{1}{\lambda_R(\mu)} + \frac{1}{2 \sqrt{\pi}} \frac{1}{\int_\Gamma d_g s} \int_0^\infty d t \int_0^\infty d u \, e^{-t^2 / 4 u} \nonumber \\
& \quad \times \left[ \iint_{\Gamma \times \Gamma} d_g \theta \, d_g \theta' \frac{e^{- \tau^{-1} \mu t}}{4 \pi u^{3/2}} K_u^{S^1} \left( \gamma(\theta), \gamma(\theta'); e^*(g) \vert_{S^1} \right) \right. \nonumber \\
& \left. \qquad - \iint_{\Gamma \times \Gamma} d_g s \, d_g s' \frac{e^{E t - m^2 u}}{\sqrt{u}} K_u \left( \gamma(s),\gamma(s'); g \right) \right] \,.
\end{align}
From this expression we easily infer that
\begin{align}
\Phi_R(\mu,\lambda_R(\mu),\tau m,\tau E,\tau^{-2} g) = \Phi_R(\tau^{-1}\mu,\lambda_R(\mu),m,E,g) \,.
\end{align}
Taking the scale-invariant derivative with respect to $\tau$ of both sides leads to the renormalization group equation for the principal operator,
\begin{align}
\left[ \tau \frac{d}{d \tau} - \beta(\lambda_R) \frac{\partial}{\partial \lambda_R} \right] \Phi_R(\mu,\lambda_R(\mu),\tau m,\tau E,\tau^{-2} g) = 0 \,.
\end{align}
In order to solve this equation, we suggest the following ansatz for the operator $\Phi_R(E)$,         
\begin{align}
\Phi_R(\mu,\lambda_R(\mu),\tau m,\tau E,\tau^{-2} g) = f(\tau)\Phi_R(\mu,\lambda_R(\tau \mu),m,E,g) \,.
\end{align}
Plugging this ansatz into the renormalization group equation gives
\begin{align}
\tau \frac{d}{d \tau} f(\tau) = 0 \,.
\end{align}
Its solution is $f(\tau)=1$ by means of the initial condition $f(\tau=1)=1$. Therefore, we obtain the desired equation for the renormalized principal operator. It suggests that the use of renormalized coupling constant at a different renormalization scale $\tau \mu$ is equivalent to the scaling of the energy, the mass, and the metric in this expression at the unchanged  renormalization scale $\mu$,
\begin{align}
\Phi_R(\mu,\lambda_R(\mu),\tau m,\tau E,\tau^{-2} g) = \Phi_R(\mu,\lambda_R(\tau \mu),m,E,g) \,.
\end{align}
We remark that here, the mass $m$ is another fixed scale in the problem, unlike a true relativistic theory where the mass itself is fixed consistently with the interactions, therefore has its own beta function. Indeed, one can go back to the exact equations for $\Phi_{ij}$ and verify this property. Therefore we have an explicit realization of how the physics at one energy scale can be understood through  the physical information at another energy scale by adjusting all other scales and couplings in the theory. 
\section{Acknowledgment}
This work is supported by Bo\u{g}azi\c{c}i University BAP Project \#:6513.

\end{document}